# Water maser emission in planetary nebulae

L. F. Miranda (1,2), O. Suárez (3), J.F. Gómez(1)

(1) Instituto de Astrofísca de Andalucía-CSIC, C/ Glorieta de la Astronomía s/n, E-18008, Granada, Spain

(2) Departamento de Física Aplicada, Facultade de Ciencias del Mar, Campus Lagoas-Marcosende, Universidade de Vigo, E-36310 Vigo, Spain (present address)

(3) UMR 6525 H. Fizeau, Université de Nice Sophia Antipolis, CNRS, OCA. Parc Valrose, 06108 Nice Cedex 2, France

**Abstract:** Planetary Nebulae (PNe) evolve from Asymptotic Giant Branch (AGB) stars after a brief post-AGB phase. Water maser emission is characteristic of oxygen-rich AGB stars, is observed in post-AGB stars and, unexpectedly, has been detected in three PNe (IRAS17347-3139, IRAS18061-2505 and IRAS19255+2123) where the physical conditions to generate water maser emission did not seem to exist. These three objects may be considered as the youngest PNe known up to date and, therefore, they are key objects to understand the formation of PNe. In addition, the existence of water maser emitting PNe allow us to study every phase in the AGB to PN transition using water maser emission which can be observed at very high spatial and spectral resolution. In this paper we review the properties of water maser emission in PNe, the existing observations of the three water maser emitting PNe and their implications in our understanding of PN formation and evolution.

## 1. INTRODUCTION

Planetary nebulae (PNe) represent the last evolutionary stage of low and intermediate mass stars (M < 8 M$_\odot$) before the white dwarf phase. They evolve from Asymptotic Giant Branch (AGB) stars after a post-AGB phase that may last from a few decades for the highest mass objects to several ten thousand years for the lowest mass ones (Blöcker 1995). Strong mass loss (up to $10^{-4}$ M$_\odot$/yr) in the AGB phase ejects the stellar envelope forming a dense expanding shell around the star. When strong mass loss stops, the star enters the post-AGB phase during which its effective temperature increases. When the effective temperature reaches ~30000 K, the shell is ionized and the star enters the PN phase.

Even though this schema is commonly accepted, the details of the transformation of an AGB star into a PN are not fully understood. This is mainly due to the dramatic change that the shell geometry suffers during the transition. Whereas AGB shells show spherical symmetry, PNe usually present axial symmetry but including complex structures like multiple shells (e.g., Guerrero et al. 2004), multiple bipolar lobes (Sahai & Trauger 1998) and highly collimated, in many cases precessing, bipolar jets (see Miranda et al. 1999 and Guerrero et al. 2009 and references therein) that can hardly be explained within an interacting winds scenario (Kwok et al. 1978; Balick 1987). The ubiquity of such complex structures in PNe has led to the suggestion that collimated outflows are the main shaping agent of PNe (Sahai & Trauger 1998). The origin of these jets is still unclear (see Balick & Frank 2002) and understanding how they form and shape PNe requires the study of the whole evolutionary sequence from AGB to PN. Water maser emission has proved to be a very useful observational tool to carry out this investigation because this emission is detected in AGB and post-AGB stars and PNe, and it can be observed at very high spatial and spectral resolution.

Water maser emission is typical of oxygen-rich AGB stars and originates in the inner regions of the spherical shell, between a few tens and few hundred AU from the star, expanding at 5-20 km/s. Post-AGB stars also show water maser emission that can exhibit very simple as well as complex features (Engels 2002; Deacon et al. 2007; Suárez et al. 2009). Particularly interesting is the case of the so called water fountains (Likkel & Morris 1988), AGB or post-AGB stars that show water maser components separated by 60 - 300 km/s, indicating high velocity outflows (e.g., Likkel & Morris 1988; Imai et al. 2002; Deacon et al. 2007; Suárez et al. 2008; see Imai 2007 and Claussen et al. 2007 for reviews). Radio interferometric observations show that water masers in water fountains trace highly collimated bipolar jets with kinematical ages of 5-100 yr only (Imai et al. 2002; Boboltz & Marvel 2005; Day et al. 2010). To date, about a dozen water fountains have been identified and they represent the first presently known manifestation of collimated mass ejection in stars evolving toward the PN phase.

The presence of water maser emission has also been confirmed in three PNe (Miranda et al. 2001; de Gregorio-Monsalvo et al. 2004; Suárez et al. 2007; Gómez et al. 2008). This result was unexpected because, in principle, the physical conditions to generate water maser emission are not fulfilled in PNe where strong mass loss does not occur and the shell is photoionized. Although the number of water maser emitting PNe (hereafter H2O-PNe) is very small, their relevance for our understanding of PN formation is crucial because H2O-PNe are the youngest PNe identified at present.

In this paper, we review the properties of water maser emission in PNe, the characteristics of H2O-PNe and their implications for our understanding of PN evolution and formation.

## 2. PROPERTIES OF WATER MASER EMISSION IN PLANETARY NEBULAE

As already mentioned, three PNe are known to exhibit water maser emission: IRAS19255+2123 (K3-35, PN G056.0+02.0; Miranda et al. 2001), IRAS17347-3139 (PN G356.8-00.0; de Gregorio-Monsalvo et al. 2004) and IRAS18061-2505 (MaC1-10, PN G005.9-02.6; Suárez et al. 2007; Gómez et al. 2008). It should be noted that water maser emission towards IRAS19255+2123 was already detected in 1984 (Engels et al. 1985) although, at that time, the association of the water masers with this emission-line nebula was not conclusively established and the true nature of the object (either young stellar object or PN) was unknown.

Before describing the properties of water masers in PNe it is important to comment about the PN nature of these objects and to describe their morphology. The latter will allow us to associate the water masers to particular nebular regions.

The PN nature of these three sources has been amply demonstrated through optical/infrared spectroscopy and radio continuum emission observations. Optical spectra of IRAS19255+2123 and IRAS18061-2505 show a typical PN spectrum (Miranda et al. 2000; Suárez et al. 2006; Chiappini et al. 2009). IRAS18061-2505 hosts a [WC8] central star (Górny & Siódmiak 2003; Górny et al. 2009), the only central star of an H2O-PN identified so far. Infrared spectra of IRAS17347-3139 shows Brα and [NeII] line emissions while optical spectra only shows relatively strong [SIII]9069 line emission (Jiménez-Esteban et al. 2006; Suárez et al. 2006). Radio continuum emission at several frequencies has been detected from the three sources (Aaquist & Kwok 1980; Aaquist 1993; Miranda et al. 2001; de Gregorio-Monsalvo 2004; Gómez et al. 2007, 2008; Tafoya et al. 2009). It is worth noting that the association of the water maser emission with these PNe is firmly established because it has been obtained with simultaneous interferometric observations of the water masers and the radio continuum emission at 1.3cm (Miranda et al. 2001; de Gregorio-Monsalvo et al. 2004; Gómez et al. 2008).

Figures 1, 2 and 3 present images of IRAS19255+2123, IRAS17347-3139 and IRAS 18061-2505, respectively. The three PNe exhibit a bipolar morphology consisting of two narrow bipolar lobes and a central region. Signs of point-symmetry in the bipolar lobes are also recognizable in the three objects, particularly, in IRAS19255+2123 in which 2cm and 3.6cm radio continuum observations show an S-shaped (precessing) bipolar jet extending from the inner nebular regions up to the bright point-symmetric knots observed at optical wavelengths (Aaquist 1993; Miranda et al. 2001; see Fig.1). We also note that IRAS17347-3139 is obscured at optical wavelengths, IRAS19255+2123 is partially obscured (its central region is optically invisible, the bipolar lobes are optically visible), while both the bipolar lobes and central region of IRAS18061-2505 are optically visible. In Figure 4 we show the water maser spectrum of IRAS18061-2505.

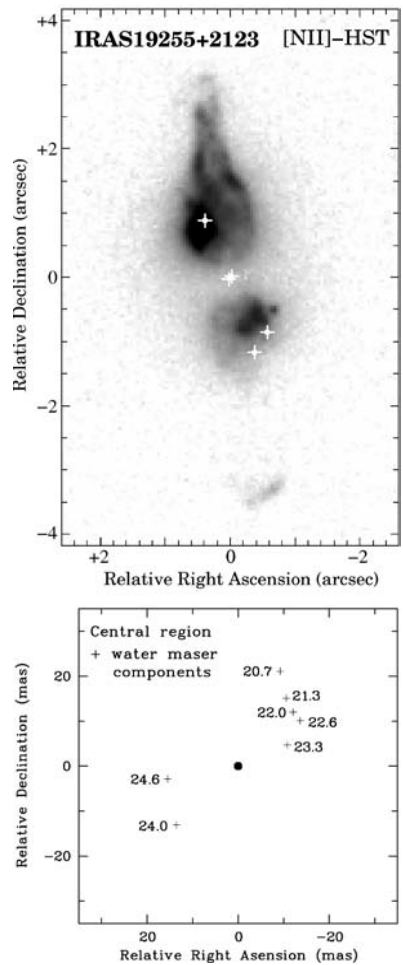

Figure 1. (top) HST image of IRAS19255+2123 in the light of [NII]. The white crosses mark the positions where water maser emission was detected in 1999. (bottom) Water maser components towards the center of the nebula as observed in 1999. The filled circle marks the position of the radio continuum emission peak at 1.3 cm. The numbers indicate radial velocities (km/s) with respect to the LSR. Note the different angular scale in the top and bottom panels.

Figs. 1 to 3 also show the location of the water maser components in the three PNe. Water masers are observed towards the central region of the nebulae and in a very small radial velocity interval. In IRAS19255+2123, water maser components are observed within 0.025" from the radio continuum emission peak and with (LSR) radial velocities between +20 and +25 km/s (Fig.1). In IRAS17347-3139, water maser components are observed within 0.25" from the center of the nebula with (LSR) radial velocities between -61 and -69 km/s (Fig.2). In IRAS18061-2505, water maser components are observed within 0.05" from the radio continuum emission peak and with (LSR) radial velocities between +57 and +61 km/s (Figs.3 and 4). In the three objects, the distribution of the water maser components is mostly perpendicular to the main nebular axis defined by the bipolar lobes. This strongly suggests that these water masers are related to dense equatorial structures (disks or toroids) in the nebulae. In IRAS19255+2123, the water masers have been modeled in terms of a slowly expanding (1.4 km/s) and rotating (3.1 km/s) disk that is oriented perpendicular to the innermost regions of the precessing jets (Uscanga et al. 2008).

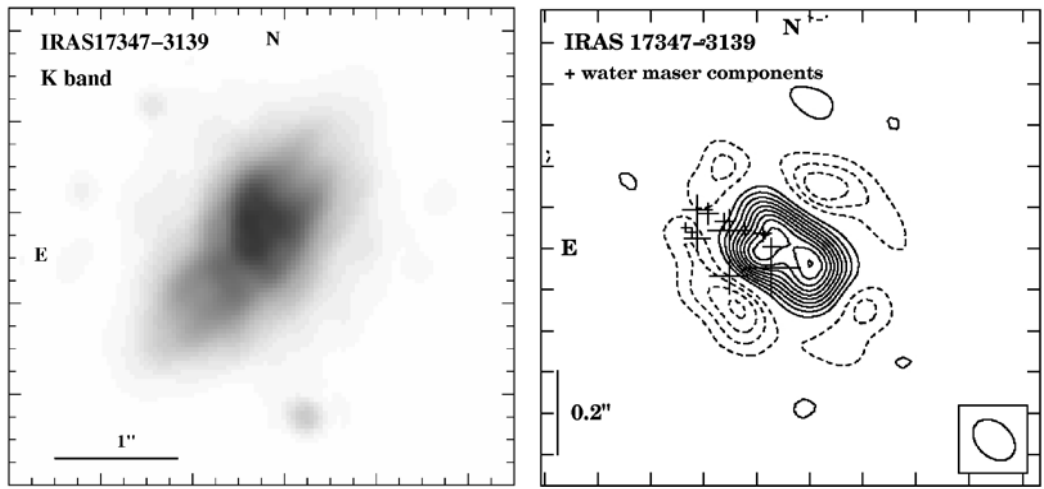

Figure 2. (left) K-band HST image of IRAS17347-3139. (right) Radio continuum emission at 0.7cm (contours) from IRAS17347-3139 and position of the water maser components (crosses) as observed in 2002 (adapted from The Astrophysical Journal, Tafoya et al. 2009). The radio continuum emission corresponds to the dark lane observed in the K image.

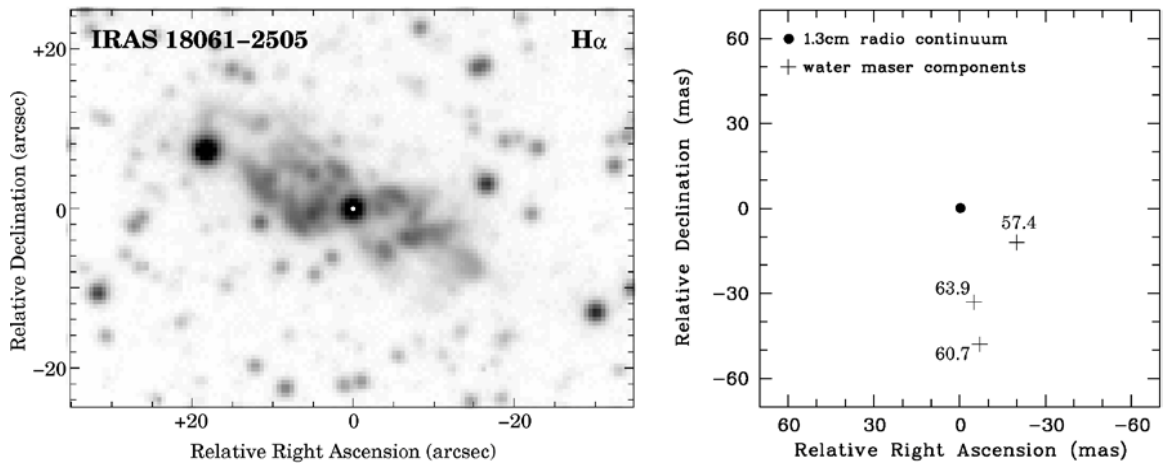

Figure 3. (left) IRAS18061-2505 in the light of Hα as observed with CAFOS at the 2.2m telescope on Calar Alto Observatory. The white square at the center indicates the region where water maser emission has been detected. (right) Positions of the three water maser components (crosses, see also Figure 4) with respect to the radio continuum peak at 1.3cm (filled circle). The numbers indicate radial velocities (km/s) with respect to the LSR. Note the different angular scale in the left and right panels.

For a distance to IRAS19255+2123 of 3.9 kpc (Tafoya et al. 2010, see below), the water masers are located at about 65 AU from the center and the radius of the water maser ring is about 80 AU (Uscanga et al. 2008). The estimated distance to IRAS17347-3139 ranges from 0.8 to 6 kpc (Gómez et al. 2007) and the water masers are located between 200 and 1500 AU for the center, respectively. In IRAS18061-2505, for a distance of 1.3 kpc to the nebula (Preite-Martínez 1998), the water masers are located at about 65 AU from the center

(Gómez et al. 2007). We note that, although water maser emission in one of the H2O-PNe has also been detected at thousands AU from the center (see below), the apparent association of the water masers with the equatorial torus of IRAS17347-3139 suggests a short distance to the object. The location of the water masers in PNe with respect to the center is similar to this of water masers in AGB stars.

In observations of 1999, water maser emission from IRAS19255+2123 was also detected at large distances (3900 and 4900 AU) from the center (Fig.1). A group of water masers was observed toward the NE with (LSR) radial velocities between +22 and +25 km/s and two groups were observed towards the SW with (LSR) radial velocities between +31 and +36 km/s, and +23 and +30 km/s. These distant water masers appear associated to the bright knots at the tips of precessing bipolar jets (Figure 1), which most probably trace the sites where the jets interact with the envelope of IRAS19255+2123 (Miranda et al. 2007). Thus, the generation of the distant water masers most probably is related to jet-envelope interaction which is able to create the suitable physical conditions to pump the water maser in distant nebular regions where such conditions do not exist in PNe.

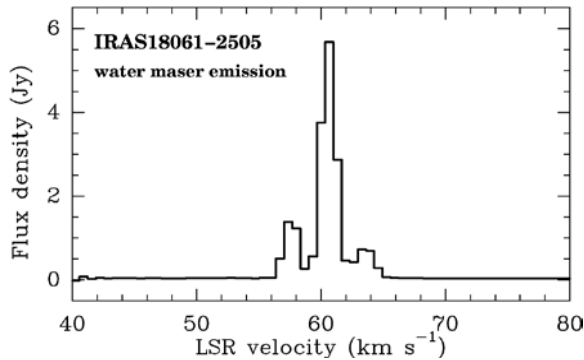

Figure 4. Water maser spectrum of IRAS18061-2505 (adapted from The Astrophysical Journal, Gómez et al. 2008).

Water maser emission in PNe displays the usual variability observed in maser emission. Strong intensity variations are observed in IRAS18061-2505 in scales of months (Suárez et al. 2007) and in IRAS19255+2123 in scales of months and years (Tafoya et al. 2010 and references therein). Positional changes of the equatorial water maser components in scales of years have also been reported in IRAS19255+2123 (Tafoya et al. 2010). The water masers in IRAS17347-3139 have been only detected during 2002 and, since then, they have disappeared (Suárez et al. 2007). The same has occurred with the distant water masers in IRAS19255+2123 which have been detected in 1999 only. Plausible explanations for the disappearance of the water masers include that the jets reach low density regions of the envelope where the conditions to generate the maser are not fulfilled (Imai 2007), the advance of an ionization front in the equatorial plane that destroys the massing regions and the creation of other massing regions farther away (Tafoya et al. 2010). However, given the extreme variability of maser emission, it cannot be ruled out that the disappeared water masers may reappear again. Monitoring of the H2O-PNe is crucial to establish possible variability patterns.

In addition to the three confirmed H2O-PNe, water maser emission has been detected towards three PN candidates using single-dish observations: IRAS 12405-6219, IRAS 15103-5754 and IRAS 16333-4807 (Suárez et al. 2009), although the association of the masers with these sources should be confirmed with radio interferometric observations and their PN nature is to be conclusively established. The water maser spectra of IRAS12405-6219 and IRAS16333-4807 show a single emission peak. In IRAS15103-5754 the water maser spectrum shows multiple components spanning a velocity range of 80 km/s that is similar to this observed in water fountains. If the PN nature of IRAS15103-5754 and its association with water maser emission are confirmed, it would be the first water fountain PN identified so far (Suárez et al. 2009).

## 3. OBSERVATIONAL PROPERTIES OF WATER MASER PLANETARY NEBULAE

Systematic observations of the three H2O-PNe as a group are very scarce and, therefore, it is difficult to establish possible similarities and differences between the three objects. As common characteristics we can mention (1) their point-symmetric bipolar morphology, strongly suggesting the action of precessing collimated outflows in their formation, and (2) a larger value of the [25]-[60] IRAS colour than water maser emitting post-AGB stars, which is also observed in the three H2O-PN candidates (Suárez et al. 2009).

IRAS19255+2123 and IRAS17347-3139 present OH maser emission which has not been detected in IRAS18061-2505 (Gómez et al. 2008). OH maser emission from IRAS19255+2123 includes emission from the four ground states at 1612, 1665, 1667 and 1720 MHz (Engels et al. 1985; te Lintel Hekkert 1991; Aaquist 1993; Miranda et al. 2001; Szymczak & Gérard 2004; Gómez et al. 2009) and emission from the first excited state at 6035 MHz (Desmurs et al. 2010). A high degree of circular polarization is observed in some features of the 1612, 1665 and 1720 MHz emissions suggesting the presence of a magnetic field (Miranda et al. 2001; Gómez et al. 2009). From the 1665 MHz emission, a value of 0.9 mG at 110 AU from the center has been obtained (Gómez et al. 2009), which is consistent with a solar-type model (see Vlemmings et al. 2005). The 1665 MHz emission is detected towards the central region of the nebula, coinciding with the regions where water maser emission is observed, suggesting the presence of a magnetized toroid (Miranda et al. 2001). In IRAS17347-3139, Zijlstra et al. (1989) first reported the presence of the 1612 MHz line. Tafoya et al. (2007) observed with the VLA the four ground state OH maser transitions but only the 1612 MHz line was detected at a radial velocity comparable to that of the water maser emission. The 1612 MHz emission did not show a particularly high circular polarization.

IRAS19255+2123 is the only known PN that shows 1720 MHz OH maser emission. This emission is close in position and radial velocity to the water masers in the center of the nebula (Gómez et al. 2009), which is consistent with the 1720 MHz emission being related to dissociation of the water molecule and both 1720 MHz and water emissions related to the same C-type shock (Sevenster & Chapman 2001). However, it is somewhat surprising that 1720 MHz emission has not been detected in the other two H2O-PNe (Tafoya et al. 2007; Gómez et al. 2008). In addition, IRAS19255+2123 is one of the two known PNe to exhibit 6035 MHz emission (Desmurs et al. 2010), the other being Vy2-2 (Jewell et al. 1985; Desmurs et al. 2010), another very young PN with 1612 MHz emission (Davis et al. 1979; Seaquist & Davis 1983; Zijlstra et al. 1989) but not water maser emission (de Gregorio-Monsalvo et al. 2004). The 6035 MHz emission in IRAS19255+2123 is also observed towards the center of the nebula and close in space and radial velocity to the 1720 MHz and water maser emissions (Desmurs et al. 2010).

The detection of OH and/or water maser emission indicates that H2O-PNe evolved from oxygen-rich AGB stars. However, the presence of emission features associated to PAHs in IRAS17347-3139 suggests the existence of considerable C-rich material in the nebulae (Jímenez-Esteban et al. 2006). A similar analysis of infrared spectra should be carried out for the other two H2O-PNe. IRAS18061-2505 shows extremely low metal abundances (Górny et al. 2009) which is not compatible with its bipolar morphology, although the N/O abundance ratio of about 1 indicates a Type I PN as many bipolar PNe. It should be noted that the [OIII] electron temperature used for the abundance calculations is unusually large (26000 K, Chiappini et al. 2009). In IRAS19255+2123 Miranda et al. (2007) found an unrealistic high (30000 K) [NII] electron temperature and evidence that the [OIII]4363 emission is contaminated by other emissions. In these circumstances, chemical abundances have not been determined for IRAS19255+2123 although a Type I PN may be suggested from its bipolar morphology and very high [NII] to Hα line intensity ratio of 5.5 (Miranda et al. 1998, 2000). For IRAS17347-3139 a Type I PN is suggested by its morphology without further available data to support this classification. High electron densities ($10^{5-6}$ cm$^{-3}$) are derived from the radio continuum flux for IRAS17347-3139 (Gómez et al. 2007) and for the inner regions and bright optical knots of IRAS19255+2123 where the [SII] line intensity ratio is at the high density limit of $10^4$ cm$^{-3}$ but it indicates values of 6000 cm$^{-3}$ in the outermost nebular regions (Miranda et al. 2000, 2007). These high values strongly contrast with the moderate electron density of 2000 cm$^{-3}$ derived from the [SII] lines in IRAS18061-2505, although it is not clear whether this is a mean value over the whole nebula or refers only to the central regions (see Górny et al. 2004). The bright optical knots of IRAS19255+2123 seem to be associated with H$_2$ emission and present a very peculiar spectrum with both low-excitation (e.g., [OI], [SII]) and high excitation emissions (e.g., HeII, [ArV], [FeVII]) arising from a relatively small region, which is compatible with shock excitation (Miranda et al. 2007).

CO and HCO$^+$ single-dish observations have been carried out towards IRAS19255+2123. The CO emission shows a broad profile that is typical of PNe (Dayal & Biening 1996; Tafoya et al. 2007) while the HCO$^+$ emission indicates a high abundance of HCO$^+$ relative to H$_2$, as compared with other young PNe (Tafoya et al. 2007). HCO$^+$ is a tracer of dense neutral regions and in IRAS19255+2123 these regions could be shielding the water molecules against the ionizing radiation from the central star (see Tafoya et al. 2007). Similar observations should be carried out towards the other two H2O-PNe to search for the possible presence and properties of these molecules. Moreover, interferometric observations of HCO$^+$ towards the three H2O-PNe are crucial to establish whether this molecule is related to the regions where the water masers arise.

Multiwavelength radio continuum observations of IRAS17347-3139 at three epochs show that the radio continuum flux density has risen (de Gregorio-Monsalvo et al. 2004; Gómez et al. 2007; Tafoya et al. 2009), which has been interpreted as due to the expansion of an ionized region with a kinematical age of 100 – 120 yr. Moreover, the turnover frequency of IRAS17347-3139 is high for PNe, which is consistent with a very young PN (see Gómez et al. 2007 and references therein). It is noticeable that the radio continuum flux density from IRAS19255+2123 does not present significant variations and that the turnover frequency is lower than that in IRAS17347-3139 (Aaquist 1993; Gómez et al. 2007). In the case of IRAS18061-2505 observations of radio continuum are available for one epoch only (Gómez et al. 2008)

Recently, the distance to IRAS19255+2123 has been determined through the annual parallax of the water maser emission to be 3.9 (+0.7/-0.5) kpc (Tafoya et al. 2010). This result is highly remarkable because the annual parallax method does not require to make assumptions on the nebular properties and, therefore, lacks the uncertainties inherent to the use of other methods to obtain distances to PNe.

## 4. EVOLUTIONARY IMPLICATIONS

Lewis (1989) and Gómez et al. (1990) proposed that the SiO, water and OH maser emissions disappear sequentially in the AGB to PN transition and estimated that water masers may last about 100 yr after (strong) AGB mass loss stops. This implies that H2O-PNe should be extremely young PNe. An estimate for the kinematical age of the three H2O-PNe can be obtained assuming an upper limit for the radius of the central ionized region as given by the distance of the water masers to the center, an expansion velocity of, say, 25 km/s, and the distances to the H2O-PNe mentioned above. With these assumptions we obtain kinematical ages of about 15 yr for IRAS19255+2123, 40-290 yr for IRAS17347-3139 and 12 yr for IRAS18061-2505. The kinematical age of IRAS17347-3139 for a distance of 6 kpc is three times higher than the time water maser are expected to survive after strong mass loss stops, suggesting that the distance to IRAS17347-3139 may be noticeable shorter than 6 kpc. In addition, a relatively short distance (about 2-3 kpc) would make compatible the kinematical age with the age of 100-120 yr estimated for IRAS17347-3139 from the increase of the radio continuum flux density. It should be noted that the very small kinematical ages obtained are mainly determined by the very small distance of the water masers to the center of the nebula and less by the assumed expansion velocity, distance to the nebula or possible projection effects, within reasonable values for these parameters. Even though the kinematical age does not necessarily represent the age of the star since photoionization starts (Schönberner et al. 2005), such very small kinematical

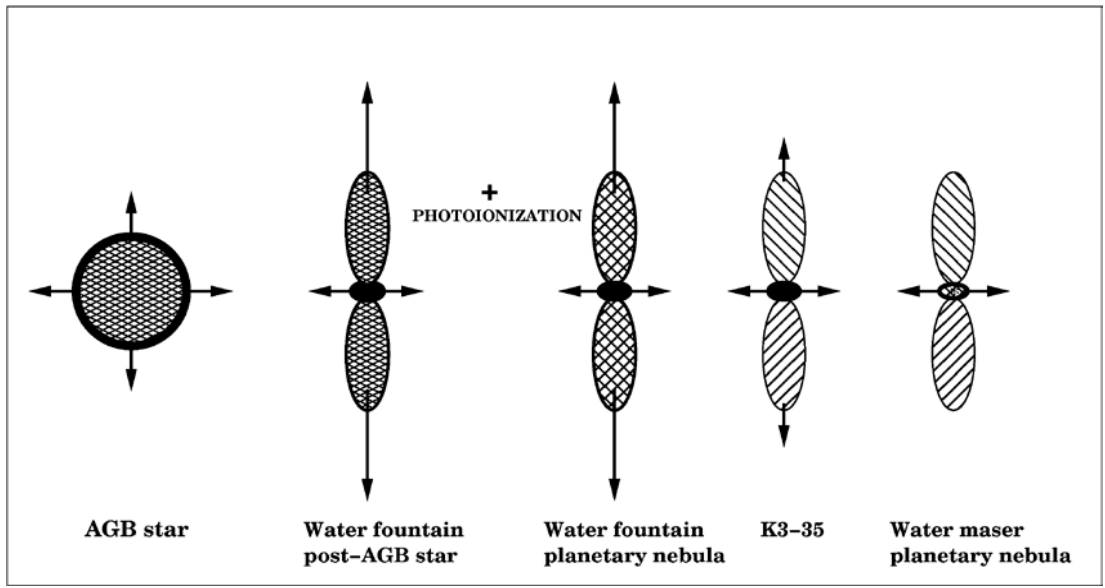

Figure 5. Possible evolutionary schema for the last evolutionary stages of 5-8 M⊙ stars based on the location in the nebula and radial velocity of water masers (following Gómez et al. 2008 and Suárez et al. 2009). The small (large) arrows indicate small (large) radial velocity of water masers. The fill pattern of the structures tries to represent the obscuration degree of the objects. The structure of a water fountain PN is speculative and based in the structures observed in the previous and next evolutionary phases. The schema does not take into account the precession of the bipolar jets observed in some water fountains and in IRAS19255+2123 (K3-35).

ages can only be obtained if the PNe indeed are extremely young.

The presence of water masers in PNe requires the coexistence of an ionized region and large amounts of dense neutral material. This suggests a very rapid evolution for the central stars of H2O-PNe. Suárez et al. (2007) suggest that H2O-PNe may derive from intermediate mass stars (5 to 8 M⊙) that traverse the post–AGB phase in < 100 yr (see Blöcker 1995). In fact, the time scale of about 100 yr for water maser disappearance may be considered as an approximate evolutionary time scale for the central stars of H2O-PNe if a photoionized region should coexist with water masers. Therefore, H2O-PNe probably are related to the final stages of the highest mass stars among low and intermediate mass stars.

The distances of the equatorial water masers in H2O-PNe to the center seem to suggest that these masers may be the remnants of AGB water masers. However, this is not necessarily true. Water masers can be excited by different winds as the star leaves the AGB phase, as a dying AGB wind or post-AGB winds (Engels 2002). In H2O-PNe it is conceivable that an incipient fast wind (or equatorial flows) drives shocks in the dense neutral equatorial regions which still present the conditions for water maser to be pumped due to the fast evolution of the central star. Very different is the case of the distant water masers in IRAS19255+2123. The enormous distance of these massing regions to the star, without precedent in any star, requires the impact of a high velocity jet to create the conditions to pump the water maser. Moreover, the association of the distant masers in IRAS19255+2123 with bipolar jets and the peculiar physical conditions of the bright optical knots can be considered as a solid evidence that jets do play a role in shaping PNe and, in particular, in the creation of point-symmetric structures.

The detection of water masers along the whole AGB to PN evolutionary sequence led Gómez et al. (2008) and Suárez et al. (2009) to propose a simple evolutionary scenario on the basis of the location in the nebula and radial velocity of the water masers. Figure 5 presents a schematic representation of this scenario that, in principle, is applicable to stars in the 5-8 M⊙ range.

The first phase corresponds to the AGB phase with low-velocity water maser emission arising from the spherical shell. The onset of anisotropic mass ejection during the late AGB or post-AGB phase marks the entrance in the water fountain phase where water masers are associated to highly collimated high velocity jets and, in some cases, to equatorial structures (Imai 2007). As already mentioned (Sect.1), only a dozen water fountains have been identified even though extensive surveys for water maser emission in post-AGB stars have been carried out (Deacon et al. 2007; Suárez et al. 2007, 2009). The small number of water fountains could be due to some bias in the observations, to the fact that (a few) water fountains may not be recognized as such if the jet moves perpendicular to the line of sight (Engels 2002; Suárez et al. 2009), or to a very rapid evolution so that the star is heavily obscured during its post-AGB phase and is difficult to identify. In this respect it is interesting to mention that Gómez et al. (2008) suggest that H2O-PNe evolve from water fountains. If so, the mass range of central stars of water fountains should be similar to that of H2O-PNe, which would explain the small number of these objects.

If evolution occurs very fast, probably for M ~ 7-8 M☉, the shell may be ionized while the water fountain jet is still active (Suárez et al. 2009). The possible water fountain PN IRAS15103-5754 would be representative of this phase. Although the structure of a water fountain PN is still unknown, it may be suggested that it will show a bipolar morphology and its water masers will be associated to bipolar jets and, perhaps, to the equatorial regions (see Fig.5).

IRAS19255+2123, the next object in the sequence, would represent the phase in which the water fountain jets are quenched, while water masers survive in the equatorial regions. We note that the distant water masers in IRAS19255+2123 show very low radial velocities (< 7 km/s, with respect to the central velocity), suggesting strong deceleration of the jets. However, these radial velocities are comparable to those of the bright optical knots (10-20 km/s, Miranda et al. 2000) and both are much lower than typical jet velocities in PNe and (polar) expansion velocities of bipolar PNe (150-300 km/s). This strongly suggests that the bright optical knots and, hence, the distant massing regions in IRAS19255+2123 move almost perpendicular to the line of sight. Proper motion measurements will certainly allow us to prove the existence of high velocities in the bright optical knots.

Finally, once the water masers associated to bipolar jets disappear, only the dense and neutral equatorial regions may preserve suitable physical conditions to generate water maser emission, as it is characteristic of the three H2O-PNe.

If we try to fit other nebular properties in this schema, the situation complicates. Several observations (obscuration at optical wavelengths, increase radio continuum flux density, turnover frequency) strongly suggest that IRAS17347-3139 is younger than IRAS19255+2123. However, the association of distant water masers with jets suggests that IRAS19255+2123 is younger than IRAS17347-3139. The non variability of the radio continuum flux in IRAS19255+2123 is not expected from an extremely young PNe. The very low metal abundances in IRAS18061-2505 are not compatible with the evolution of a 5-8 M☉ star. Moreover, while IRAS17347-3139 and IRAS19255+2123 exhibit signs of extreme youth, there is no observational evidence to date in IRAS18061-2505 that reveals an extreme youth, except for the presence of water maser emission. Most probably the detailed post-AGB evolution will critically depend on the stellar mass at the end of AGB, which will determine the phases in Fig.5 and the associated time scales the nebula will go through. However, despite the varied phenomenology observed in these three objects (and, in general, in post-AGB stars and PNe), it is remarkable that two properties of the water maser emission (location in the nebula and radial velocity) are able to arrange, at least qualitatively, a very simple evolutionary sequence for evolved stars.

The presence of 1720 MHz OH emission in IRAS 19255+2123 and not in the other two H2O-PNe indicates the existence of different physical conditions at scales below 100 AU. The detection of the central star of IRAS18061-2505 at optical wavelengths makes it surprising the presence of water masers at 50 mas (about 65 AU) from the central star, suggesting the existence of complex structures below that radius. Because H2O-PNe are making their entrance in the PN phase, they are ideal objects to study the phenomenology associated to PN formation, which, as suggested by the observations, is highly complex at scales < 100 AU or even much smaller. Multiwavelength observations of H2O-PNe capable to resolve these regions will certainly provide information to make a substantial progress in our understanding of PN formation.

## 5. CONCLUSIONS

Water maser emission has been confirmed to exist in three PNe (named H2O-PNe): IRAS19255+2123, IRAS17347-3139 and IRAS18061-2505. The water masers are observed towards the central regions of the nebula, presumably related to small (radius < 100 AU) equatorial disks or toroids, and are detected in a very small radial velocity interval. In IRAS19255+2123 water maser emission has been detected in one epoch at enormous distances (3900 – 4900 AU) from the center at the tips of bipolar jets in the object. In addition, water maser emission has been detected towards three PN candidates, one of them being a water fountain.

The observational data indicate that H2O-PNe are the youngest PNe identified so far and probably related to the last evolutionary stages of 5 - 8 M☉ stars. The three present a point-symmetric bipolar morphology indicating that precessing bipolar jets are involved in their formation. Moreover, the distant masers in IRAS19255+2123 provide conclusive evidence that jets are involved in the shaping of PNe.

H2O-PNe allow us to study the whole evolutionary AGB to PN sequence using water maser emission. Despite the complex and varied phenomenology associated to post-AGB stars and PNe, it is remarkable that the location in the nebula and the radial velocity of the water masers are able to arrange the objects in a simple (qualitative) evolutionary scenario.

The confirmation of the three H2O-PN candidates as bona fide H2O-PNe and systematic observations of the true H2O-PNe are crucial to establish common properties as well as differences among the objects. Given that H2O-PNe are just entering the PN phase and that most data point out that the innermost regions (< 100 AU) of H2O-PNe are the sites of complex physical processes, spatially and spectrally resolved observations of these regions are crucial to clarify these processes and, therefore, to understand PN formation.

## Acknowledgments

We are very grateful to the editors of this book for the invitation to write this review. This work has been supported partially by Spanish Ministerio de Ciencia e Innovación grants AYA2008-06189-C03 (JFG, OS) and AYA2008-01934 (LFM, OS) (both co-funded with FEDER funds), and grants INCITE09E1R312096ES and

## References

Aaquist, O.B., 1993, A&A, 267, 260
Aaquist, O.B., Kwok, S., 1990, A&AS84, 229
Balick, B., 1987, AJ, 94, 671
Balick, B., Frank, A., 2002, ARA&A, 40, 439
Boboltz, D.A., Marvel, K.B., 2005, ApJ, 627, L45
Chiappini, C., Górny, S.K., Stasińska, G., Barbuy, B., 2009, A&A, 494, 591
Claussen, M.J., Sahai, R., Morris, M., 2007, Asymmetrical Planetary Nebulae IV, eds. R.M.L. Corradi, A. Manchado, N. Soker, published on line http://www.iac.es/project/apn4, article #48
Davis, L.E., Seaquist, E.R., Purton, C.R., 1979, ApJ, 230, 434
Day, F.M., Pihlström, Y.M., Claussen, M.J., Sahai, R., 2010, ApJ, 713, 986
Dayal, A, Bieging, J.H., ApJ, 472, 703
De Gregorio-Monsalvo, I., Gómez, Y., Anglada, G., et al., 2004, ApJ, 601, 921
Deacon, R.M., Chapman, J.M., Green, A.J., Sevenster, M.N., 2007, ApJ, 658, 1096
Desmurs, J.-F., Baudry, A., Sivagnanam, P., et al., 2010, A&A, 520, 45
Engels, D., 2002, A&A, 388, 252
Engels, D., Schmid-Burgk, J., Walmsley, C.M., Wimberg, A., 1985, A&A, 148, 344
Gómez, J.F., de Gregorio-Monsalvo, I., Lovell, J.E.J., et al., MNRAS, 364, 738
Gómez, J.F., Suárez, O., Gómez, Y., et al., 2008, AJ, 135, 2074
Gomez, Y., Tafoya, D., Anglada, G., et al., 2009, ApJ, 695, 930
Górny, S.K., Siódmiak, N., 2003, in Planetary Nebulae: Their Evolution and Role in the Universe, ed. S. Kwok, M. Dopita, R. Sutherland, IAU Symp. 209, 43
Górny, S.K., Chiappini, C., Stasińska, G., Cuisinier, F., 2009, A&A, 500, 1089
Górny, S.K., Stasińska, G., Escudero, A.V., Costa, R.D.D., 2004, A&A, 427, 231
Guerrero, M. A., Chu, Y.-H., Miranda, L. F., 2004, AJ, 128, 1694
Guerrero,M.A., Miranda, L.F., Riera, A., et al., 2008, ApJ, 683, 272
Imai, H., 2007, in Astrophysical Masers and their Environments, ed. W. Baan, & J. Chapman (Cambridge University Press), IAU Symp. 242, 279
Imai, H., Obara, K., Diamond, P.J., et al., 2002, Nat, 417, 829
Jiménez-Esteban, F.M., Perea-Calderón, J.V., Suárez, O., et al., 2006, in Planetary Nebulae in our Galaxy and Beyond, ed. M.J. Barlow & R.H. Méndez, IAU Symp. 234, 437
Kwok, S., Purton, C.R., Fitzgerald, P.M., 1978, ApJ, 219, L125
Lewis, B.M., 1989, ApJ, 338, 234
Likkel, L., Morris, M., 1988, ApJ, 329, 914
Miranda, L.F., Fernández, M., Alcalá, J.M., et al., 2000, MNRAS, 311, 748
Miranda, L.F., Gómez, Y., Anglada, G., Torrelles, J.M., 2001, Nat, 414, 284
Miranda, L.F., Guerrero, M.A., Torrelles, J.M., 1999, AJ, 117, 1421
Miranda, L.F., Luridiana, V., Guerrero, M.A., et al., 2007, Asymmetrical Planetary Nebulae IV, eds. R.M.L. Corradi, A. Manchado, N. Soker, published on line http://www.iac.es/project/apn4, article #34
Miranda, L.F., Torrelles, J.M., Guerrero, M.A., et al., 1998, MNRAS, 298, 243
Preite-Martínez, A., 1988, A&AS, 76, 317
Sahai, R., Trauger, J.T., 1998, AJ, 116, 1357
Schönberner, D., Jacob, R., Steffen, M., 2005, A&A, 441, 573
Seaquist, E.R., Davis, L.E., 1983, ApJ, 274, 659
Sevenster, M.N., Chapman, J.M., 2001, ApJ, 546, L119
Suárez, O., García-Lario, P., Manchado, A., et al., 2006, A&A, 458, 173
Suárez, O., Gómez, J.F., Miranda, L.F., 2008, ApJ, 689, 430
Suárez, O., Gómez, J.F., Miranda, L.F., et al., 2009, A&A, 505, 217
Suárez, O., Gómez, J.F., Morata, O., 2007. A&A. 467, 1085
Tafoya, D., Gómez, Y., Anglada, G., et al., 2007, AJ, 133, 364
Tafoya, D., Gómez. Y., Patel, N.A., et al., 2009, ApJ, 691, 611
Tafoya, D., Imai, H., Gómez, Y., et al., 2010, PASJ, submitted
te Lintel Hekkert, P., 1991, A&A, 248, 209
Uscanga, L., Gómez, Y., Raga, A.C., et al., 2008, MNRAS, 390, 1127
Vlemmings, W.H.T., van Lagevelde, H.J., Diamond, P.J., 2005; A&A, 434, 1029
Zijlstra, A.A., te Lintel Hekkert, P., Pottasch, S. R., et al., 1989, A6A, 217,